\documentclass[
]{ceurart}

\usepackage{listings}
\usepackage{dot2texi}
\usepackage{tikz}
\usetikzlibrary{shapes,arrows}

\usepackage{algorithm}
\usepackage{algpseudocode}

\begin{document}

\copyrightyear{2022}
\copyrightclause{Copyright for this paper by its authors.
  Use permitted under Creative Commons License Attribution 4.0
  International (CC BY 4.0).}

\conference{CLEF 2022: Conference and Labs of the Evaluation Forum, 
    September 5--8, 2022, Bologna, Italy}

\title{Motif Mining and Unsupervised Representation Learning for BirdCLEF 2022}

\author[1]{Anthony Miyaguchi}[
orcid=0000-0002-9165-8718,
email=acmiyaguchi@gatech.edu,
url=https://acmiyaguchi.me,
]
\author[1]{Jiangyue Yu}[
email=jyu478@gatech.edu
]
\author[1]{Bryan Cheungvivatpant}[
email=bcheung0@gatech.edu
]
\author[1]{Dakota Dudley}[
orcid=0000-0001-8352-0391,
email=dakotadudley13@gatech.edu,
]
\author[1]{Aniketh Swain}[
email=aswain9@gatech.edu
]

\address[1]{Georgia Institute of Technology, North Ave NW, Atlanta, GA 30332}

\begin{abstract}
    We build a classification model for the BirdCLEF 2022 challenge using unsupervised methods. We implement an unsupervised representation of the training dataset using a triplet loss on spectrogram representation of audio motifs. Our best model performs with a score of 0.48 on the public leaderboard.
\end{abstract}


\maketitle

\section{Introduction}

The BirdCLEF 2022 challenge involves identifying species of native Hawaiian birds from soundscape recordings. The training dataset comprises 14.8 thousand recordings totaling over 190 hours from Xeno-canto \cite{xeno-canto} of varying length and quality. The specific task involves predicting the presence of a list of 29 bird species in 5-second non-overlapping windows of 1-minute soundscape recordings. The recordings are Ogg Vorbis audio files at a sample rate of 32khz. The Xeno-canto training examples are labeled by species but not at 5-second intervals per the challenge's primary task. Therefore, they may contain both ambient noise and birdcalls from other species.

We focus our efforts on unsupervised methods to tackle the lack of concrete labels for the multi-label classification problem. First, we experiment with motif mining algorithms to identify windows of audio that contain birdcalls as an unsupervised process for generating labels. The motif mining process also provides metadata used to train downstream models. Additionally, instead of training a classification model directly from training examples, we choose to build an embedding that captures similarities between birdcalls across all training examples. We finally train classification models using the embedding model to reduce the dimensionality of the original data.

\section{Motif Mining}

We utilize a motif mining algorithm called SiMPle \cite{silva2018fast} to extract the segments of the audio clip that best represent common patterns in the clip. SiMPle provides two primary operations to compute summary data structures called the matrix profile and profile index: a self-join that computes the similarity within a track and a join that computes the similarity between two tracks. The procedure involves converting the raw audio track into a spectrogram to capture the frequency components of the audio over a sliding window in time. Then we apply SiMPle to compute the matrix profile and profile index, providing the distance to the nearest neighbor and the index of the nearest neighbor in the set of windows in the track. A motif is a window of an audio track with the lowest distance to all other windows in the join operation as given by the matrix profile. The indices of the matrix profile's minimum and maximum distance values are the motifs and discord, respectively. 

\begin{figure}[h]
\includegraphics[width=\textwidth]{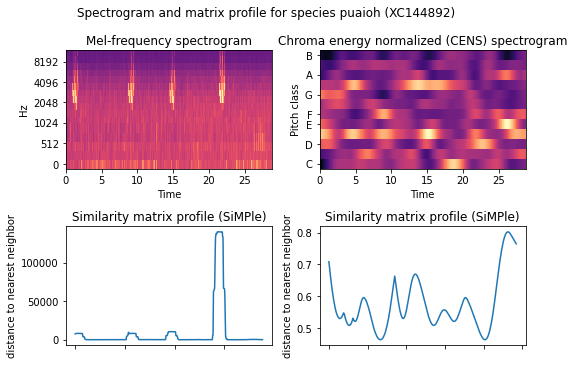}
\caption{
    Spectrograms show frequency components of audio transformed via STFT. We apply SiMPLe to obtain a matrix profile that summarizes the distance to the nearest neighbor for all time-slices in the spectrogram. Spectrogram parameterization affects the quality of the matrix profile summary.
}
\label{fig:spectrogram}
\end{figure}

We hypothesize that the motifs and discords capture the audio clip's salient features, i.e., bird calls. Much like how SiMPle can identify the chorus of a song via the motif, it may be possible to extract birdcalls as a motif from the training examples. We transform the training dataset by extracting the matrix profile and profile index information from each track via a self-join and using the 5-second motifs as a soft label for birdcalls in each species. We utilize the self-join index profile during the training procedure of the birdcall embedding. We also test the feasibility of using a matrix profile join as a feature in the classifier for the main BirdCLEF task by directly utilizing the resulting join against a random set of motifs extracted from the training dataset.

\section{Birdcall Embedding}

An embedding maps data from one space into a lower-dimensional space while maintaining relative distances between mapped data. We experiment with creating a birdcall embedding using a modified Tile2Vec model \cite{jean2019tile2vec}. The Tile2Vec model utilizes a sampling procedure on spatially distributed data and a triplet loss to learn a lower-dimensional representation of Earth imagery tiles that retains semantic similarity between tiles. The triplet loss takes advantage of the triangle inequality by forming triplets between anchor, neighbor, and distant tiles. We borrow the loss for triplets $(t_a, t_n, t_d)$ with a margin $m$ where $f_{\theta}$ maps audio data to a $d$-dimensional vector of real numbers using a model with parameters $\theta$. 

\begin{equation}
    L(t_a, t_n, t_d) = \left[
        ||f_{\theta}(t_a) - f_{\theta}(t_n)||_2
        - ||f_{\theta}(t_a) - f_{\theta}(t_d)||_2
        +m
    \right]_+
\end{equation}

We hypothesize that we can learn an embedding by forming triplets using windows of audio that have similar spectral qualities. We utilize the SiMPLe index of the audio spectrogram to determine the neighborhood of any given window of audio within a track. The Tile2Vec model is a modified ResNet-18 convolutional neural network -- we make further modifications to generate spectrograms in the correct shape as the first layer in the model.

\section{Experiments}

\subsection{Motif Mining Details}

We begin by extracting the motifs and discords of the entire training dataset. However, this process is computationally intensive because it requires streaming the entire audio dataset, computing the spectrogram via STFT, and computing the matrix profile. Therefore, we pre-compute the matrix profile and index profile for offline processing. We use librosa \cite{librosa} for audio loading and transformation and implement a NumPy implementation of the SiMPle algorithm to facilitate usage in our processing pipeline. \footnote{Implementation at \
\href{https://github.com/acmiyaguchi/simple-fast-python}{github.com/acmiyaguchi/simple-fast-python}}

We use chroma energy normalized (CEN) and Mel-scaled spectrograms. We generate CEN spectrograms at ten samples per second using the default parameters in librosa, with a 50 sample window for motif mining purposes and the Mel-scaled spectrograms using an FFT window of 2048, a hop length of 80 samples, and 16 Mel bands with a 400 sample window. 

\begin{table}[]
\begin{tabular}{|l|l|l|l|l|l|}
\hline
spectrogram format & transform time & shape      & size   & SiMPle window & self-join \\ \hline
Mel-scaled         & 131ms          & (16, 7932) & 126912 & 400           & 3.3s      \\ \hline
CEN                & 287ms          & (12, 292)  & 3504   & 50            & 8.85ms    \\ \hline
\end{tabular}
\caption{Statistics from processing track XC144892 which is 28.8 seconds long. Each operation is run 10 times. We note that our parameterization of the Mel-scaled spectrogram results in data that takes several orders of magnitude longer to process than the CEN spectrograms.}
\label{tab:spectrogram}
\end{table}

As per figure \ref{fig:spectrogram}, we note that the observed quality of a spectrogram and the resulting matrix profile varies depending on the algorithm and parameters used. We can clearly distinguish the chirps in the high-frequency ranges in the Mel spectrogram, with obvious discords occurring in the same time intervals. We are presented with a lower resolution spectrogram with CEN because it is aligned by chroma instead, which does not delineate between the bird call and the background noise.

We initially performed our motif mining using the CEN parameters, following closely with the SiMPLe cover song extraction experiment. We found that the lower resolution of the CEN spectrogram representation allowed us to process all training examples and store the primary representation on disk, as per table \ref{tab:spectrogram}. Although noisy, the representation did extract bird calls as the primary motif, although lack of domain knowledge prevented us from quantifying the quality of these labels. As a result, some extracted motifs are background noise or human voices. 

\subsection{Embedding Clustering Quality}

\begin{figure}[hbt!]
\centering
\begin{tikzpicture}[>=latex',scale=0.7]
\begin{dot2tex}[dot,tikz,codeonly]
    digraph G {
        rankdir=LR;
        subgraph cluster_birdclef {
            label="birdclef-2022"
            train[label="training audio", shape=rect]
        }

        subgraph cluster_motif {
            label="motif mining"
            pi[label="profile index", shape=rect]
        }        
        subgraph cluster_embedding {
            label="embedding"
            subgraph cluster_dataloader {
                label="dataloader"
                triplets[shape=rect]
            }
            subgraph cluster_model {
                label="model"
                mels[label="spectrogram", shape=rect]
                tilenet[shape=rect]
            }
            triplets -> mels
            mels -> tilenet
            
        }
        train -> pi
        pi -> triplets
        train -> triplets
    }
\end{dot2tex}
\end{tikzpicture}
\caption{Flow diagram of the constituent pieces of the birdcall embedding.}
\end{figure}

We reuse the open-sourced Tile2Vec model and add a Mel spectrogram layer via nnAudio \cite{nn-audio}. We use an FFT window of 4096 with a calculated hop length matching the height of 128 Mel bands. We choose these parameters to mimic the 128x128 pixel images used to train the source model.

We built two separate data loaders for this task — the first data loader pre-computed triplets from the original Ogg Vorbis files into NumPy array files. We only use the extracted motifs during the data loading process. In addition, we attempt to adjust for the skewed species distribution by oversampling underrepresented species. A second data loader streamed all audio computing triplets using the pre-computed SiMPle indices for each track. We break each track into windows of 5-seconds and assign the nearest neighbor using the profile index. Next, we create several queues, each containing window pairs from tracks of different species of birds. The data loader pops pairs from each queue to form a mini-batch, after which triplets are formed by randomly assigning the third element from an element within the mini-batch. Finally, we augment the audio tracks once formed into triplets. 

\begin{algorithm}
    \caption{Sampling triplets from audio using precomputed SiMPle index}
    \begin{algorithmic}
        \Require $D$ is the set of audio tracks as raw samples
        \Require $f_s$ is the sampling rate of the audio in Hz
        \Require $s$ is the size of the audio window in seconds
        \Require $I$ is the SiMPle index
        \Ensure $T$ is the set of triplets formed by all windows
        \Ensure $|T| = |P|$
        \State Initialize pairs $P = \{\}$
        \For{$y \in D$}
            \State $w \gets \text{window}(y, f_s, s)$
            \Comment{$w$ is an array of audio windows each of length $s \times f_s$}
            \For{$i \gets 1, \text{length}(w)$}
                \State $P \gets P \cup (y, w[i], w[I[i]])$
                \Comment{Use the SiMPle index to get the nearest neighbor}
            \EndFor
        \EndFor
        \State $T \gets \left\{ 
            (x_a, x_n, y_a) | 
                \,
                \forall (x, x_a, x_n) \in P, 
                \exists((y, y_a, y_n) \in P 
                ,
                \, x \neq y
        \right\}$
    \end{algorithmic}
\end{algorithm}

The first approach is untenable for an online training procedure because it naively generated motif triplets which required upwards of $3N$ disk reads. In addition, we noticed that running the data loading procedure online would cause underutilization between the CPU and GPU due to the loading bottleneck. By storing $5e5$ NumPy array triplets to disk, we could fully utilize the GPU at the cost of disk space, going from the original 6GB of audio data to 135GB for the pre-computed triplet data. The second approach required no additional memory because it took on an iterable approach to creating triplets. It has a different distributional semantic from the first since it includes all audio windows instead of just the track motifs. We also do not address class imbalances because online undersampling is difficult. We need further preprocessing to determine the total number and locations of birdcalls per class.

We apply a random gain of $[-20, 20]$ dB, Gaussian random noise between $[-5, 40]$ dB, and pitch shift between $[-4, 4]$ semitones using the audiomentations library \cite{audiomentations}. We note that augmentation can become a bottleneck in the data loading process. Instead of applying it to the entire triplet, it suffices to apply it to just the motif pairs before triplet formation. We consider torch-audiomentations but had poor performance due to difficulties applying the correct device to torch tensors inside the data loader.

We validate the embedding learned on the iterable data loader by training a classifier on motifs transformed into the embedding space. We chose a subset of three species: \texttt{brnowl}, \texttt{skylar}, and \texttt{houfin}. These species are common and have several hundred audio tracks available. We generate a training dataset comprised of the motif of each track and cross-validate a logistic regression model using model accuracy on species accuracy. We sample $k=300$ motifs for ten models and report the mean and standard deviation of the accuracy on embedding models that vary in the output size.

\begin{figure}[h]
    \centering
    \includegraphics[width=0.75\textwidth]{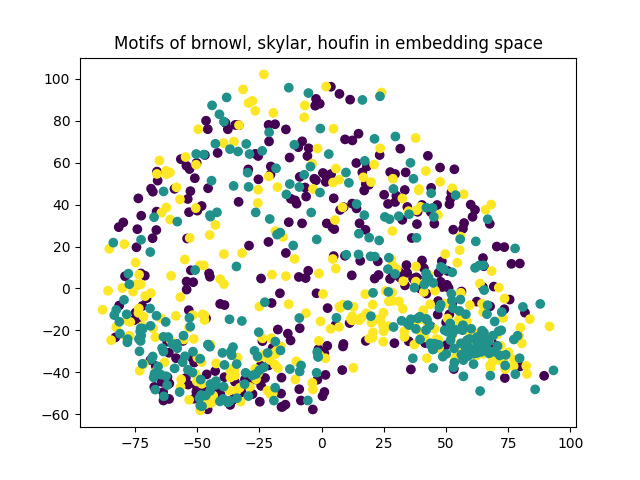}
    \caption{A scatter plot of a random set of motifs (n=300) drawn from three species of birds. The motifs are truncated, padded, and transformed into the embedding space. We plot the top two principle components found by PCA. }
    \label{fig:my_label}
\end{figure}

We note a slight difference between accuracy in models of equivalent parameterization when the output dimension of the model increases from 256 to 512 from $0.53$ to $0.54$.

\begin{table}[h]
\begin{tabular}{|l|l|}
\hline
dimension & accuracy (n=10) \\ \hline
128       & 0.528 (0.031)   \\ \hline
256       & 0.529 (0.019)   \\ \hline
512       & 0.544 (0.048)   \\ \hline
\end{tabular}
\caption{Accuracy table of a logistic regression model trained on the embedding using motif mined from three species of birds. }
\label{tab:emb-acc}
\end{table}

\subsection{Classifier Performance}

We train a classifier using the embedding model as a preprocessing step. The first classifier trains a gradient-boosted decision tree (GBDT) via LightGBM \cite{ke2017lightgbm}. It uses a MultiOutputClassifer to create a model per task to suit the multi-label classification task using the default parameters of the LightGBM classifier. Next, we apply a similar set of augmentation to the embedding model, with a random gain of $[-20, 20]$ dB, pitch shift between $[-4, 4]$ semitones, time shift between $[-0.1, 0.1]\%$ of the window's length, and colored noise between $[-3, 30]$ dB and $[-2, 2]$ frequency power decay using the PyTorch-audiomentations library \cite{pytorch-audiomentations}. We also train a multi-layer perceptron (MLP) to perform multi-label classification directly using a binary cross-entropy loss. The loss function allows us to implement mixup \cite{zhang2017mixup} during data loading mini-batches as an augmentation process. 

We experimented with using matrix profile join data as a feature in our classification model. We draw 64 motif samples from a set of training motifs and run each audio window in the testing task through a matrix profile against the training sample. We use the concatenation of the matrix profile join's min, median, and max as an additional feature in the model. Unfortunately, we found that the additional CPU overhead to run SiMPle far outweighed any marginal performance benefit to the model and was omitted in subsequent model iterations.

\begin{table}[h]
\begin{tabular}{|l|l|}
\hline
classifier     & macro F1-score \\ \hline
GBDT           & 0.0177         \\ \hline
MLP            & 0.0151         \\ \hline
\end{tabular}
\caption{A comparison of the two classifiers' performance on a subset of the training audio data, using macro F1-score as metric. We limit the scored data to the first ten tracks of each species in the task.}
\label{tab:f1-score}
\end{table}

We compute the macro F1-score in table \ref{tab:f1-score} to compare our best GBDT model against our best MLP model. We note that the GBDT edges slightly over the MLP, but performance for both classifiers is poor. We often see a lack of positive predictions for a class, which leads to a score of 0. We also assume that all audio windows contain birdcalls to simplify the calculations, so these scores understate the actual performance of the models. However, we found that the general performance is still poor in the BirdCLEF task, hovering around 0.48 on the public leaderboard during the competition.

\section{Discussion}

In a small set of shorter tracks (30 seconds or less), we consistently found that the discord tends to be a birdcall instead of the motif. It might be more productive to standardize a mining procedure on shorter tracks instead of using the whole, variable-length tracks to take advantage of this observation. We can build a no-call detector using the matrix profile as a feature if we can consistently detect positions of bird calls through discords or motifs. In addition, we can achieve higher throughput on parallelization by reducing data skew in the distribution of training example lengths. Finally, both spectrogram transformations via nnAudio and the SiMPle algorithm can be written in PyTorch to increase the performance of the motif mining algorithm.

We found that the birdcall motif triplet embedding performed poorly for downstream prediction tasks. We hypothesize that the spectrogram parameters during motif mining cause the embedding to rest on a representation ill-suited for classification. One modification to the embedding triplet procedure would be to form pairs of anchors and neighbors from overlapping audio windows. These pairs would be similar due to their distance in time. As the experimental procedure before, triplets are formed by randomization in mini-batches. A sliding window triplet procedure would enable a comparison of the validity of the learned motif triplet embedding.

\section{Conclusions}

We implemented and experimented with the SiMPle motif mining algorithm, a modified Tile2Vec embedding model, and several multi-label classification models.\footnote{Implementation at \
\href{https://github.com/acmiyaguchi/birdclef-2022}{github.com/acmiyaguchi/birdclef-2022}} 
Our performance on the leaderboard was underwhelming, but we solved many engineering problems throughout the challenge with potential improvements for future BirdCLEF challenges.

\begin{acknowledgments}
Thanks to the Data Science @ Georgia Tech (DS@GT) officers for organizing and publicizing recruitment for the DS@GT Kaggle competition team.
\end{acknowledgments}

\bibliography{dsgt-birdclef-2022}
\appendix
\end{document}